\documentclass[12pt]{article}
\usepackage{graphicx} 
\usepackage[export]{adjustbox} 
\title{AdS/CFT without holography: \\
A hidden dimension on the CFT side \\
and implications for black-hole entropy}
\author{Hrvoje Nikoli\'c \\
Theoretical Physics Division, Rudjer Bo\v{s}kovi\'{c} Institute, \\
P.O.B. 180, HR-10002 Zagreb, Croatia \\
{\normalsize e-mail: hnikolic@irb.hr} \\
\makebox[1in]{} \\
}
\date{\today}
\begin{document}
\maketitle
\begin{abstract}
We propose a new non-holographic formulation of AdS/CFT correspondence, according to which
quantum gravity on AdS and its dual non-gravitational field theory
both live in the same number $D$ of dimensions. The field theory, however, appears
$(D-1)$-dimensional because the interactions do not propagate in one of the dimensions.
The $D$-dimensional action for the field theory can be identified with the sum over 
$(D-1)$-dimensional actions with all possible values $\Lambda$ of the UV cutoff,
so that the extra hidden dimension can be identified with $\Lambda$.
Since there are no interactions in the extra dimension, most of the
practical results of standard holographic AdS/CFT correspondence transcribe
to non-holographic AdS/CFT without any changes.
However, the implications on black-hole entropy change significantly.
The maximal black-hole entropy now scales with volume, while the Bekenstein-Hawking entropy
is interpreted as the minimal possible black-hole entropy. In this way,
the non-holographic AdS/CFT correspondence offers a simple resolution of the black-hole information paradox,  
consistent with a recently proposed gravitational crystal.
\end{abstract}
\vspace*{0.5cm}

\section{Introduction}

\subsection{Main ideas}
\label{SEC1.1}

It is by now a well established fact that there is a close relationship between quantum gravity
in $D$-dimensions and gauge theory without gravity in $(D-1)$-dimensions. 
(See e.g. \cite{maldacena,gkp,witten} for the original formulation, \cite{aharony} for an early review,
\cite{natsuume} for very readable introductory lectures, and \cite{ammon} for a comprehensive monograph.)
However, it is one thing to claim that two theories are related, and another
to claim that they are completely equivalent to each other. While the former is a well established 
fact, the latter is still an unproven conjecture. 

In this paper we present some arguments that the conjecture, as usually formulated, 
might not be completely correct. As a possible alternative, we propose that the usual formulation of the conjecture
should be replaced with a modified one, according to which the gravity theory in $D$ dimensions 
is equivalent to a gauge theory without gravity which also lives in $D$ dimensions.
However, the gauge theory is not an ordinary gauge theory. It is a somewhat peculiar $D$-dimensional 
theory which, in many respects, looks like an ordinary gauge theory in $(D-1)$-dimensions.
Even though this field theory lives in $D$ dimensions, the interactions propagate
only in $(D-1)$ dimensions. This makes one of the dimensions hidden, explaining why the usual
$(D-1)$-dimensional view of the gauge theory works so well. 

\begin{figure*}[t]
\includegraphics[width=11cm,center]{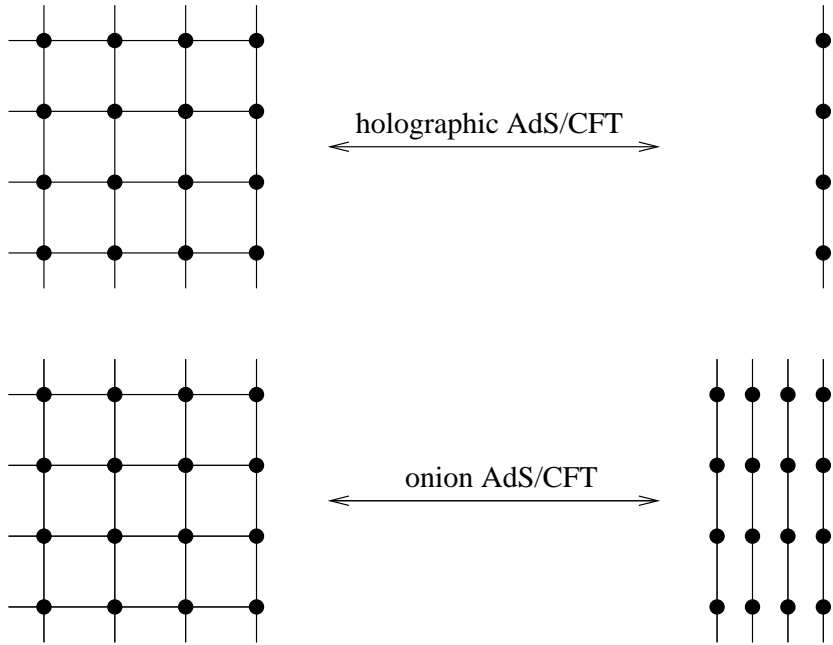}
\caption{\label{fig1}
A lattice visualization of two different versions of AdS/CFT duality. The links represent the 
interactions between the degrees of freedom represented by the dots. The AdS theory on the left is supposed
to be dual to the CFT theory on the right. The holographic CFT has one dimension less 
than the AdS, while the onion CFT has the same number of dimensions as AdS. Consequently, 
the onion CFT has more degrees of freedom than the holographic CFT.}
\end{figure*}

Since the standard formulation of the conjecture claims that a $D$-dimensional theory is equivalent
to a $(D-1)$-dimensional theory, and since it is best understood in the case when
the gravitational $D$-dimensional theory is formulated on the AdS-background in which case the gauge theory is a
conformal field theory, we refer to the
standard duality conjecture as holographic AdS/CFT duality. By contrast, 
in our reformulation, the spacetime in which gauge theory lives can be visualized as a family of 
many $(D-1)$-dimensional layers which do not interact with each other, so we refer to 
it as {\em onion} AdS/CFT duality. The two different versions of AdS/CFT duality are depicted in Fig.~\ref{fig1}.

If the gauge theory looks $(D-1)$-dimensional because
there are no interactions in the one extra dimension, then what is gained by the extra dimension?
The answer is - {\em entropy}. Even though there are no interactions between different
layers, all these layers may contribute to the total entropy. Even though 
the contributions from different layers cannot be measured by a hypothetic CFT observer living
at a single layer, the contributions from different layers are still there. Moreover, 
the contributions from different layers can be measured by the observer living in the dual theory,
namely in AdS. The immediate consequence of this is that the measurable entropy in quantum gravity
should scale with {\em volume}, rather than area.

Of course, the general idea that a theory in $D$ dimensions is dual to another theory in $D$ dimensions,
and consequently, that entropy should scale with volume in both theories, 
certainly does not sound very surprising. However, when one of the theories contains gravity,
there seems to be a serious problem with such an idea. The entropy scaling with volume 
seems to be in contradiction with the Bekenstein-Hawking entropy 
\begin{equation}\label{BH}
 S_{\rm BH}\equiv\frac{A}{4} ,
\end{equation}
which scales with the area $A$ (expressed in natural units with unit Planck length). 
Several independent arguments strongly suggest that (\ref{BH}) is equal to the entropy of the black hole.
Indeed, the Bekenstein-Hawking entropy of black holes is one of the main arguments
for validity of the holographic version of AdS/CFT duality. So how can non-holographic onion 
AdS/CFT duality be compatible with our knowledge on black hole entropy?

Since the problem of black-hole entropy seems, at first look, to be the main problem for our proposal
of non-holographic AdS/CFT duality, we shall discuss this problem first, in Sec.~\ref{SEC2}.
We shall argue that the black-hole entropy $S_{\rm bh}$ satisfies the inequality
\begin{equation}
 S_{\rm bh}\geq S_{\rm BH},
\end{equation}
where the equality is only valid initially, when the entropy contained in Hawking radiation can be neglected. 

The standard holographic form of AdS/CFT correspondence has been very successful in various applications,
such as those in QCD \cite{QCD1,QCD2,QCD3}, theory of fluids \cite{fluid1,fluid2,fluid3,fluid4}, 
theory of superconductivity \cite{supercond1,supercond2,supercond3}, and entanglement entropy \cite{ent1,ent2,ent3}.
How can such a success of the holographic version of AdS/CFT correspondence be explained by
the non-holographic version? Our answer is that these applications have not much to do 
with the full {\em equivalence} between the gauge theory and gravity. We make this point
sharp in Sec.~\ref{SEC3}, where we point out that the GKPW relation \cite{gkp,witten}, 
which is the best known and most important
equation in AdS/CFT correspondence, is {\em not} an evidence for the conjecture that the bulk theory in AdS
is completely {\em equivalent} to the CFT theory on the boundary. We support this claim
by deriving a generalized GKPW relation,
valid for theories for which it is completely clear that they are not equivalent to each other.

To familiarize with the idea of physical theories with an onion structure, in Sec.~\ref{SEC4} we study  
some simple toy models with such a structure. Then in Sec.~\ref{SEC5} we propose that the CFT theory
(dual to quantum gravity in AdS) has onion structure with a hidden dimension identified with 
the UV cutoff $\Lambda$. 

Finally, the conclusions are drawn in Sec.~\ref{SEC6}.

\subsection{Notes on terminology}
\label{SEC1.2}

In the literature, the expressions ``AdS/CFT correspondence'' and ``AdS/CFT duality'' are usually used as
synonyms. We find it confusing, so we shall distinguish these two concepts in the following way.
By ``AdS/CFT correspondence'' we shall merely mean the existence of a non-trivial {\em relation} between the two theories.
By ``AdS/CFT duality'' we shall mean the full {\em equivalence} of the two theories. In this sense,
the claim of AdS/CFT duality is much stronger than the claim of AdS/CFT correspondence.

The AdS/CFT correspondence, or more generally the bulk/boundary correspondence, 
can be used in two different ways. One way to use it is to exploit the bulk theory 
to gain knowledge about the boundary theory. The other way is the opposite, 
to exploit the boundary theory to gain knowledge about the bulk theory. 
The theory one is really interested about in a given context will be referred to as {\em target theory}.
The other theory, which is used to gain knowledge about the target theory,
will be referred to as {\em tool theory}. In most applications of bulk/boundary correspondence,
the target theory is a boundary theory, while the tool theory is a bulk theory.
In the next section (Sec.~\ref{SEC2}), however, our target theory will be the bulk theory. 
   
\section{Black-hole entropy without holography}
\label{SEC2}

It is widely believed that black-hole entropy $S_{\rm bh}$ is equal to
\begin{equation}\label{bh=BH}
 S_{\rm bh}= S_{\rm BH},
\end{equation}   
where $S_{\rm BH}$ is the Bekenstein-Hawking entropy (\ref{BH}). The three main arguments for this belief are
\begin{enumerate}
 \item Black-hole entropy in classical and semi-classical gravity.
 \item Black-hole entropy in string theory.
 \item Black-hole entropy according to AdS/CFT correspondence.
\end{enumerate}
Let us analyze each of these arguments separately.

\subsection{Black-hole entropy in classical and semi-classical gravity}

The laws of classical black-hole evolution take a form analogous to the laws of thermodynamics 
\cite{bekenstein,bch},
with surface gravity at the horizon being analogous to temperature, 
and surface area being analogous to a quantity proportional to the entropy. 
Further insight comes from semi-classical gravity, where matter is quantized in a fixed gravitational
background of a classical black hole \cite{BD}. As originally shown by Hawking \cite{hawking},
the black hole produces radiation with a thermal spectrum, at the temperature determined by 
the global black-hole properties such as mass, charge and angular momentum. For instance, for the
Schwarzschild black hole of mass $M$ in 4 dimensions, the Hawking temperature turns out to be
\begin{equation}\label{T}
T=\frac{1}{8\pi M} .
\end{equation}
The temperature fixes the constant of proportionality between area and entropy, leading to the universal result 
\begin{equation}\label{bh=BHsemicl}
 S_{\rm bh}^{\, \rm semiclas}= S_{\rm BH},
\end{equation}
independent on mass, charge, angular momentum, or number of dimensions.
Since this is a semi-classical result, it should be trusted when the conditions for validity
of semi-classical approximation are satisfied. In other words, (\ref{bh=BHsemicl})
should be the correct black-hole entropy when the effects of quantum gravity can be neglected. 

Naively, one might assume that effects of quantum gravity can be neglected whenever the black hole is macroscopic.
However, such an assumption leads to a serious trouble known as the {\em black-hole information paradox} 
\cite{gid,har,pres,pag,gid2,str,math1,math2,hoss,harlow,fabbri}.
The semi-classical calculation shows that Hawking radiation is in the mixed state, while the standard laws 
of quantum time evolution do not allow evolution from a pure to a mixed state (see however 
\cite{nik1,nik2,nik3,nik4} for 
alternative quantum laws of time evolution). The mixed nature of Hawking radiation strongly suggests that 
radiation must be correlated with some additional degrees of freedom, but then these additional degrees of freedom
carry entanglement entropy equal to the entropy of radiation. 
The question is - where do these additional degrees live?

If they live {\em outside} of the black hole, then the full black hole entropy can be equal to (\ref{bh=BHsemicl}),
but the problem is that 
it is not compatible with the semi-classical theory of Hawking radiation.
Namely, the semi-classical theory predicts that outside Hawking radiation is entangled with negative-energy quanta
which enter the black-hole {\em interior}. In this sense the semi-classical theory of Hawking radiation
predicts that the additional degrees live {\em inside} the black hole, in contradiction with the semi-classical
relation (\ref{bh=BHsemicl}). In any case, the semi-classical theory cannot be self-consistent even for 
macroscopic black holes. One way or another, the effects of quantum gravity must be significant even 
when the black hole is large.

How can quantum effects be significant at the macroscopic scale?
As emphasized in \cite{nik-crystal}, there is actually nothing unusual about that.
This is especially true if the quantum effects do not involve quantum {\em coherence}
of a large number of degrees of freedom. For instance, the existence of an ordinary crystal is a well known example 
of a macroscopic quantum object,
which was used in \cite{nik-crystal} to propose that the additional black-hole degrees of freedom
might be the degrees of a gravitational crystal. The entropy of a crystal naturally scales with volume, not unlike 
the entropy of the lattice in the lower-left corner in Fig.~\ref{fig1}. 
In such a scenario, the final state of Hawking radiation is a large remnant 
with a mass and size much above the Planck scale \cite{nik-crystal}.

To conclude, if (\ref{bh=BHsemicl}) is the full black-hole entropy of the initial state before 
the start of Hawking radiation,
the requirement that the semi-classical process of Hawking radiation should be unitary suggests that
(\ref{bh=BHsemicl}) should be improved by the inequality 
\begin{equation}\label{bh>BH2}
 S_{\rm bh}\geq S_{\rm BH},
\end{equation}
where the equality is only valid in the initial state. The inequality (\ref{bh>BH2}) violates
the Bekenstein bound, which we discuss in more detail in Sec.~\ref{SECBB}

\subsection{Black-hole entropy in string theory}
\label{SEC2.2}

Unlike semi-classical gravity, the string theory is a fully quantum gravity theory. 
One of the greatest successes of string theory is the discovery \cite{strominger-vafa,horowitz,peet} 
of fully quantum black-hole 
states with entropy equal to
\begin{equation}\label{bh=BHstring}
 S_{\rm bh}^{\, \rm string}= S_{\rm BH} .
\end{equation}
At first sight, it seems to be in contradiction with (\ref{bh>BH2}), which allows black-hole entropy 
to be even larger than $S_{\rm BH}$. 

Nevertheless, there is no contradiction! The entropy on the left-hand side of (\ref{bh=BHstring}) is entropy 
of a {\em specific state} (typically a certain highly symmetrical configuration of branes)
with given values of mass, charge and angular momentum. If we denote this specific state by $s^{(0)}$, 
then Eq.~(\ref{bh=BHstring})
can be written more correctly as
\begin{equation}\label{bh=BH0}
 S_{\rm bh}^{(0)}\equiv S_{\rm bh}[s^{(0)}] = S_{\rm BH} .
\end{equation} 
The explicit identification of this state in string theory does not imply that there is no another state $s^{(1)}$
with the same values of mass, charge and angular momentum, and yet with an even 
larger entropy $S_{\rm bh}^{(1)}>S_{\rm bh}^{(0)}$.
Moreover, there may exist a whole family of states $s^{(n)}$ with entropies
\begin{equation}
 S_{\rm bh}^{(n)}>S_{\rm bh}^{(n-1)}.
\end{equation}
As far as we are aware
such higher entropy states have not yet been found in string theory, but there is no proof that they do not exist.
Our non-holographic AdS/CFT correspondence makes a non-trivial prediction that such states should exist
in the full theory of quantum gravity, so it would be interesting to search for such states in non-perturbative
string theory or M-theory. The explicit search for such states is, however, beyond the scope of the present paper,
so we leave it for the future investigations. Let us only note that,
perhaps, such states could be related to the monster states studied in \cite{hsu}.

Our prediction of higher-entropy states has interesting implications on black-hole thermodynamics.
The entropy of a statistical state with given values of conserved quantities such as mass, charge and angular momentum
is maximal in the thermal equilibrium. This implies that black hole with the Bekenstein-Hawking 
entropy (\ref{bh=BH0}) is {\em not} in the thermal equilibrium. Yet, such a black hole has a temperature,
given e.g.~by (\ref{T}). How can a black hole have temperature without being in the thermal equilibrium?

The answer is in fact very simple. The mentioned temperature is the temperature of Hawking radiation, 
and Hawking radiation is produced at the horizon. This implies that the mentioned temperature is the 
temperature of the horizon. However, a black hole in thermal equilibrium must have the same temperature
{\em everywhere} within the black hole, not only at its boundary at the horizon. This strongly suggests that
the lowest possible entropy in (\ref{bh=BH0}) is really the entropy of the horizon (which explains why it is
proportional to the horizon area), while the interior degrees of freedom carry no entropy
in the state satisfying (\ref{bh=BH0}). For the same reason the interior degrees of freedom do not have
temperature (or have zero temperature $T=0$), implying that the black hole interior is not in thermal
equilibrium with its boundary at finite temperature.

A state out of thermal equilibrium tends to evolve towards the full thermal equilibrium.
How such a thermalization takes place within the black hole? A full answer would require a full
theory of quantum gravity, but a partial answer is already given by the semi-classical theory.
The Hawking radiation is produced in pairs \cite{hawking}, with one member of the pair going outside,
and another member traveling to the black-hole interior. In this way, the black-hole horizon at temperature $T$
thermalizes not only with the black-hole exterior, but also with the black-hole interior.
This process of thermalization activates the internal degrees of freedom inside the black hole,
which increases the total entropy of the black hole in agreement with (\ref{bh>BH2}). 

\subsection{Black-hole entropy according to AdS/CFT duality}

Since CFT is clearly a unitary theory, the AdS/CFT duality predicts that quantum gravity is also a unitary theory.
But we need to distinguish the prediction of the usual holographic AdS/CFT duality 
from those of our non-holographic AdS/CFT duality. 

Unfortunately, as discussed in \cite{mathur-dialog}, 
the holographic version of the duality, by itself, does not say 
how exactly the unitarity realizes in the case of Hawking radiation and what exactly 
is wrong with the semi-classical scenario which predicts the violation of unitarity. 
 
Can the non-holographic version of AdS/CFT duality do better? Yes! Unlike the holographic version,
the non-holographic version is perfectly consistent with (\ref{bh>BH2}), and consequently
with the scenario described in Sec.~\ref{SEC2.2}. So let us continue with the description 
that we interrupted at the end of Sec.~\ref{SEC2.2}.

The process of thermalization of the black-hole
interior is finished when all the interior degrees of freedom are thermalized. According to our non-holographic 
AdS/CFT duality the number of interior degrees of freedom scales with volume $V$, so the corresponding maximal
black hole entropy is expected to be of the order
\begin{equation}\label{SV}
 S_{\rm bh}^{\, \rm max} \sim V .
\end{equation}
For instance, for a large black hole with radius $R$ in $D=4$ spacetime dimensions we have $S_{\rm bh}^{\rm max}\sim R^3$,
which is much larger than the corresponding Bekenstein-Hawking entropy $S_{\rm BH}\sim R^2$. 

What happens with the black hole when it reaches this maximal possible entropy? 
The gravitational-crystal model \cite{nik-crystal} suggests that at that point 
classical general relativity ceases to be a good 
approximation at the boundary of the hole. In fact, it is even misleading to call it a ``black hole'' 
because the boundary cannot longer be interpreted as a classical horizon.
At this point the production of further radiation at the boundary either stops or becomes more like radiation
produced by an ordinary non-gravitational object at finite temperature. In either case,
the existence of such a macroscopic remnant with entropy of the order of (\ref{SV}) resolves 
the black-hole information paradox.

\subsection{What is wrong with the Bekenstein bound?}
\label{SECBB}

According to the strong version of the Bekenstein bound \cite{bek-bound,susskind-lindesay}
(see however \cite{smolin-bek-bound} for weaker versions),
the entropy of a region bounded
by the area $A$ cannot be larger than the Bekenstein-Hawking entropy (\ref{BH}). 
The inequality (\ref{bh>BH2}) clearly violates the Bekenstein bound. If (\ref{bh>BH2})
is correct, then what exactly is wrong with the Bekenstein bound?

First, the derivation of the Bekenstein bound is based on the {\em assumption} \cite{susskind-lindesay}
that the black-hole
entropy is equal to the Bekenstein-Hawking entropy. If that assumption can be violated
(as we argued to be the case), then the logic used to derive the Bekenstein bound cannot longer be applied.
Second, the derivation of the Bekenstein bound contains a tacit assumption that matter (which contains entropy
and enters the region bounded by $A$) carries {\em positive} energy. That assumption is clearly violated by Hawking radiation,
because Hawking radiation is associated with {\em negative} energy entering the black hole \cite{BD}.   

Indeed, suppose that the black-hole entropy is equal to the Bekenstein-Hawking entropy at some initial time $t_0$.
Then, according to the semi-classical theory, even the creation of a single Hawking pair at $t>t_0$ will
necessarily violate the Bekenstein bound. This, in fact, will happen for two reasons.
First, because the entropy of the interior will 
{\em increase} due to absorption of one member of the entangled Hawking pair.   
Second, because the area of the black hole will {\em decrease} due to absorption of negative energy.

Thus we see that the Bekenstein bound is very difficult to reconcile with the 
existence of Hawking radiation. To resolve the information paradox associated with the existence of 
Hawking radiation, abandoning the strong version of the Bekenstein bound looks like a very natural way out.
But note that a weaker version of the Bekenstein bound, valid only in processes which do not involve absorption of 
negative energy, is still consistent with our findings.

\subsection{Conclusion}

In this section we have shown that non-holographic AdS/CFT duality is compatible with 
current knowledge about black-hole entropy. Moreover, when dealing with the black-hole
information paradox, our arguments suggest that the non-holographic version works much better
than the usual holographic one. This motivates us to further explore arguments for validity
of the non-holographic version of AdS/CFT duality, which is the subject of the rest of the paper.

\section{The GKPW relation and its generalization}
\label{SEC3}

\subsection{The ordinary GKPW relation}

Probably the most important and best known equation of AdS/CFT correspondence, 
or more generally bulk/boundary correspondence, is the GKPW relation \cite{gkp,witten}
\begin{equation}\label{3.1}
 Z_{\rm bulk}[\Phi_{bound}=J]=Z_{\rm bound}[J] .
\end{equation}
Here $Z_{\rm bulk}$ is the partition function of quantum gravity in the bulk,
which (in the real-time version) can be written in a slightly more explicit form as
\begin{equation}\label{3.2}
 Z_{\rm bulk}=\int{\cal D}\Phi\, e^{iA_{\rm bulk}[\Phi]} ,
\end{equation}
where $A_{\rm bulk}[\Phi]$ is the action of quantum gravity expressed as a functional of generic fields $\Phi$.
The notation $Z_{\rm bulk}[\Phi_{bound}=J]$ denotes that the boundary fields $\Phi_{bound}$ are not integrated 
out in (\ref{3.2}), but are fixed to some arbitrary values $J$. The right-hand side of (\ref{3.1}) is the 
generating functional for the boundary theory, which can be written in a more explicit form as
\begin{equation}\label{3.3}
 Z_{\rm bound}[J]=\int{\cal D}\phi \, e^{i\left( A_{\rm bound}[\phi]+\int d^{D-1}x\, J(x){\cal O}(x) \right)} ,
\end{equation}
where $A_{\rm bound}[\phi]=\int d^{D-1}x\,{\cal L}[\phi(x)]$ is the action of boundary theory 
expressed as a functional of generic fields $\phi$. The fields ${\cal O}(x)$ are either the 
fundamental boundary fields $\phi(x)$ themselves, or some composite fields constructed from $\phi(x)$.
The generating functional can be used to calculate the correlation functions in the boundary theory.
The full correlation functions are given by
\begin{equation}\label{3.4}
 \langle {\cal O}(x_1)\cdots {\cal O}(x_k) \rangle = \left. \frac{1}{i^k} 
\frac{\delta^k Z_{\rm bound}[J]}{\delta J(x_1)\cdots \delta J(x_k)} \right|_{J=0} ,
\end{equation}
but physically more relevant are their connected parts
\begin{equation}\label{3.5}
 \langle {\cal O}(x_1)\cdots {\cal O}(x_k) \rangle_{\rm connect} = \left. \frac{1}{i^k} 
\frac{\delta^k W_{\rm bound}[J]}{\delta J(x_1)\cdots \delta J(x_k)} \right|_{J=0} ,
\end{equation} 
where $Z_{\rm bound}[J]=e^{iW_{\rm bound}[J]}$.

Eq.~(\ref{3.1}) is particularly useful in the regime in which the quantum theory on the left-hand side 
can be approximated by the corresponding classical theory. In this case only the classical field configurations
have a significant contribution to the integral in (\ref{3.2}), so, up to an irrelevant normalization factor, 
(\ref{3.1}) can be approximated by
\begin{equation}\label{3.6}
\left. e^{iA_{\rm bulk}[\Phi]} \right|_{\Phi_{bound}=J} = Z_{\rm bound}[J] .
\end{equation}
The normalization factor is irrelevant because it does not contribute to (\ref{3.5}).

Note that equation (\ref{3.6}) relates a classical theory on the left to a quantum theory on the right.
In general, it is very difficult to calculate the quantum functional integral in (\ref{3.3}), while it is
usually much simpler to calculate the classical on-shell action on the left-hand side of (\ref{3.6}).
This is why Eq.~(\ref{3.6}) is very useful for practical calculations when the target theory 
(the theory one is really interested about) is the boundary theory. Indeed, 
most practical applications of the bulk/boundary correspondence are applications to a non-gravitational
target theory on the boundary, including QCD \cite{QCD1,QCD2,QCD3}, theory of fluids \cite{fluid1,fluid2,fluid3,fluid4}, 
and theory of superconductivity \cite{supercond1,supercond2,supercond3}.
On the other hand, Eqs.~(\ref{3.6}) and (\ref{3.1}) have not been so much useful to gain a new knowledge
about the bulk theory.

After this brief overview of well known stuff about the GKPW relation, now we come to the point.
The two most important equations of AdS/CFT correspondence, namely Eq.~(\ref{3.1}) and its approximative form
(\ref{3.6}), {\em are not expressions of AdS/CFT duality!} Namely, even though these expressions establish
a relation between the two theories, we claim that the validity of these expressions is {\em not an evidence that 
the bulk theory is fully equivalent to the boundary theory}. 

To demonstrate that validity of (\ref{3.1}), by itself, is not an evidence of duality,
our strategy is to show that an equation of the form similar to (\ref{3.1}) may be valid 
for {\em another} two theories, for which it is totally clear that they are not equivalent to each other.
In fact, to make the demonstration more powerful, we shall show that an equation similar
to (\ref{3.1}) is valid for {\em any} quantum system on the left-hand side (playing the role of a ``bulk'') 
and its restriction to its {\em arbitrary} subsystem on the right-hand side (playing the role of a ``boundary'').
Such a generalized GKPW relation is the subject of the next subsection.  

\subsection{Generalized GKPW relation}
 
Quite generally speaking, it should not be too surprising that information about the boundary 
can be encoded in a bulk theory.
After all, the boundary is a part of the bulk, so everything what is present at the boundary (and possibly much more)
should also be present in the bulk. Or more generally, all information about a subsystem (and possibly much more)
should also be encoded in the whole system.

Consider an arbitrary physical system with canonical configuration variables
\begin{equation}
 Q=\{ Q_1,\ldots,Q_N\} .
\end{equation}
For simplicity we work with a finite number $N$ of degrees of freedom, but our results can also be generalized
to systems with infinite number of degrees of freedom. Let us make an arbitrary split of the system into two
subsystems
\begin{equation}
 Q=(b,B),
\end{equation}
where 
\begin{equation}
 b=\{ Q_1,\ldots,Q_n\} , \;\;\; B=\{ Q_{n+1},\ldots,Q_N\} ,
\end{equation}
and $n<N$. The $Q$-system will play a role of a ``bulk'',
the $b$-subsystem will play a role of a ``boundary'',  and the $B$-subsystem will
correspond to all the ``bulk'' variables which are not present in the ``boundary'' subsystem.

Now assume that the action
\begin{equation}\label{AQ} 
A[Q]=A[b,B] 
\end{equation}
is given. We shall not write down the action explicitly because 
we are making a general analysis, meaning that the action can be arbitrary.
The quantum partition function associated with this action is
\begin{equation}\label{3.2g}
 Z=\int{\cal D}Q \, e^{iA[Q]}=\int{\cal D}b \int{\cal D}B  \, e^{iA[b,B]} .
\end{equation}
Formally we can write ${\cal D}Q=\prod_{t} \prod_{j=1}^N dQ_j(t)$, and similarly for ${\cal D}b$ and ${\cal D}B$.
The partition function (\ref{3.2g}) is analogous to (\ref{3.2}). Likewise, we can introduce the quantity
\begin{equation}\label{g1}
 Z[b]=\int{\cal D}B\, e^{iA[b,B]} ,
\end{equation}
which is analogous to the left-hand side of (\ref{3.1}).

Now let us introduce the Fourier transform $\tilde{Z}[\varphi]$ of $Z[b]$. It is given by 
\begin{equation}
\tilde{Z}[\varphi] = \int{\cal D}'b\, Z[b] e^{-ib\varphi} ,
\end{equation}
where
\begin{equation}
 b\varphi\equiv \int dt \sum_{j=1}^n b_j(t)\varphi_j(t) ,
\end{equation}
and the prime reminds us that the measure ${\cal D}'b=\prod_{t} \prod_{j=1}^n db_j(t)/(2\pi)$
contains the appropriate $2\pi$-factors.
The inverse Fourier transform is then
\begin{equation}\label{g2}
 Z[b]=\int{\cal D}\varphi\, \tilde{Z}[\varphi] e^{ib\varphi} .
\end{equation}
This can be written as 
\begin{equation}\label{g4}
 Z[b]=\int{\cal D}\varphi\, e^{i(\tilde{A}[\varphi]+b\varphi)} ,
\end{equation}
where $\tilde{A}[\varphi]$ is {\em defined} by the relation
\begin{equation}\label{g3}
 \tilde{Z}[\varphi] \equiv e^{i\tilde{A}[\varphi]} . 
\end{equation}
Eq.~(\ref{g4}) is analogous to (\ref{3.3}). If we interpret $\tilde{A}[\varphi]$ as the action
for the configuration variables $\varphi=\{\varphi_1,\ldots,\varphi_n \}$,
then (\ref{g4}) is the generating functional for the corresponding correlation functions.
For instance, we have
\begin{equation}\label{g5}
 \langle \varphi_1(t_1)\cdots \varphi_k(t_k) \rangle = \left. \frac{1}{i^k} 
\frac{\delta^k Z[b]}{\delta b_1(t_1)\cdots \delta b_k(t_k)} \right|_{b=0} ,
\end{equation} 
which is analogous to (\ref{3.4}). Finally, from (\ref{g1}) and (\ref{g4}) we see that we can write
\begin{equation}\label{g1-4}
 \int{\cal D}B\, e^{iA[b,B]} = \int{\cal D}\varphi\, e^{i(\tilde{A}[\varphi]+b\varphi)} .
\end{equation}

Eq.~(\ref{g1-4}) is analogous to the GKPW relation (\ref{3.1}). Since we derived (\ref{g1-4})
under very general assumptions, we refer to (\ref{g1-4}) as {\em generalized GKPW relation}.
In particular, Eq.~(\ref{g1-4}) relates a theory with $N$ degrees of freedom on the left
to a theory with $n<N$ degrees of freedom on the right. Since 
the theory on the left-hand side of (\ref{g1-4}) can be {\em any} theory with $N$ degrees of freedom, 
and since the subsystem on the right-hand side of (\ref{g1-4}) can be an {\em arbitrary} subsystem
with an {\em arbitrary} number $n<N$ degrees of freedom, it is clear that, in general, {\em the theories
defined by the actions $A[Q]$ and $\tilde{A}[\varphi]$ are not equivalent}.
Even though $A[Q]$ and $\tilde{A}[\varphi]$ are closely related,
{\em $\tilde{A}[\varphi]$ is not dual to $A[Q]$}.

\subsection{Discussion}

The moral of the analysis above is the following: Since, in general, the generalized GKPW relation (\ref{g1-4}) does not
indicate an equivalence between two theories, {\em we should not take the ordinary GKPW relation (\ref{3.1})
as evidence for the equivalence between the bulk theory and the corresponding boundary theory}.
Using the terminology of Sec.~\ref{SEC1.2},
Eq.~(\ref{3.1}) certainly expresses a kind of bulk/boundary {\em correspondence}, but it does not express
a bulk/boundary {\em duality}. 

It does not, however, mean that the content of the ordinary GKPW relation (\ref{3.1}) is trivial.
In general, the action $\tilde{A}[\varphi]$ defined by (\ref{g3}) can be very complicated,
even when it originates from a relatively simple action $A[Q]$ in (\ref{AQ}).
The action for an open subsystem can be much more complicated than the action for the whole closed system,
especially if the subsystem is not chosen in a particularly clever way.
On the other hand, the ordinary GKPW relation (\ref{3.1}) claims that the boundary theory is a very 
nice local field theory, like a gauge theory with conformal invariance. 
Such a nice local field theory may be of interest in its own right as a target theory, 
which is the reason why the ordinary GKPW relation (\ref{3.1}) is typically much more
useful than the generalized GKPW relation (\ref{g1-4}). 

\section{Toy models with onion structure}
\label{SEC4}

In Sec.~\ref{SEC1.1} we introduced the idea that a $D$-dimensional system with interactions 
propagating in $D$ dimensions can be dual to an onion system, i.e. a $D$-dimensional system with interactions
propagating in $(D-1)$ dimensions (see Fig.~\ref{fig1}).
In this section we describe some simple explicit examples of systems with such a structure.

\subsection{Onion with two degrees of freedom}

The simplest system with an onion-like structure has two degrees of freedom.
For instance, consider the Lagrangian
\begin{equation}\label{L1}
 L=\frac{\dot{X}^2_1}{2} + \frac{\dot{X}^2_2}{2} -\frac{\omega^2}{2}(X_1-X_2)^2 ,
\end{equation}
describing the dynamics of two configuration variables $X_1$ and $X_2$ coupled 
via the harmonic potential proportional to $(X_1-X_2)^2$.
Since the two degrees of freedom are coupled we can depict such a system as 
\ \mbox{\raise0.0ex\hbox{$-$}\kern-1.2em $\bullet\;\,\bullet$}, which looks like a piece 
of lattice on the left-hand side of Fig.~\ref{fig1}.

Now let us describe the same system with the new configuration variables
\begin{equation}
 Y_1=X_1-X_2 , \;\;\; Y_2=\frac{X_1+X_2}{2}.
\end{equation}
In these new variables, the Lagrangian (\ref{L1}) takes the decoupled form
\begin{equation}\label{L2}
 L=\frac{\dot{Y}^2_1}{4} + \dot{Y}^2_2 -\frac{\omega^2}{2}Y_1^2 .
\end{equation}
Since now there is no coupling between the two degrees of freedom
we can depict it as $\bullet\;\bullet$, which looks like a piece 
of onion on the right-hand side of Fig.~\ref{fig1}.

Since there is no coupling between $Y_1$ and $Y_2$, these two degrees of freedom evolve
independently and one degree of freedom cannot ``know'' about the existence of the other. 
From the point of view of any of them, the system looks as if there was only one 
degree of freedom. Yet, the true number of degrees of freedom is two,
and they can ``know'' about each other in the dual picture described by $X_1$ and $X_2$. 

The moral of this simple example is that a system in which different degrees of freedom
interact with each other may be dual to a decoupled system in which  different degrees of freedom
do not interact with each other. In our case, \ \mbox{\raise0.0ex\hbox{$-$}\kern-1.2em $\bullet\;\,\bullet$}
is dual to $\bullet\;\bullet$.

The example above can also be generalized to a larger discrete system with a full lattice/onion 
structure as in the lower half of Fig.~\ref{fig1}. However, we shall not pursue this explicitly.
Instead, in the next example we shall study a continuous system. 

\subsection{Unparticle onion}

Unparticle field theory is a peculiar scale invariant field theory,
the manifestation of which, under certain conditions, may look like a non-integer
number of invisible particles \cite{georgi}.
Such a field theory is characterized by a continuous spectrum of mass \cite{step,kras,gaete,nik-unpart}.
The relation between unparticles and AdS/CFT correspondence has been studied in \cite{step,cacci},
but here we shall study a simpler case of flat Minkowski geometry.

Consider the scale invariant action
\begin{equation}\label{unpA}
 A=-\frac{1}{2}\int d^4x\, \eta^{\mu\nu} \partial_{\mu}\phi\,\partial_{\nu}\phi ,
\end{equation}
where $\eta^{\mu\nu}$ is the Minkowski metric with the signature $(-+++)$. 
It describes the dynamics of the field 
\begin{equation}
\phi(x)\equiv\phi(x^0,x^1,x^2,x^3)\equiv\phi(x^{(3)},z) ,
 \end{equation} 
where $x^{(3)}\equiv\{x^0,x^1,x^2\}$, $ x^3\equiv z$.
The action (\ref{unpA}) is usually called ``free'', but such a name is misleading
because the terms proportional to $-(\partial_i\phi)^2$ for $i=1,2,3$ are really 
interactions between the neighboring points in the 3-dimensional space.
For instance, by writing 
\begin{equation}
\partial_z\phi(z)= \frac{\phi(z+dz)-\phi(z)}{dz} ,
\end{equation}
one recognizes that $-(\partial_z\phi)^2$ is essentially a continuous version 
of the harmonic interaction in (\ref{L1}).

Now let us introduce the partial Fourier transform
\begin{equation}
 \tilde{\phi}(x^{(3)},k_z)=\int dz\,\phi(x^{(3)},z)\, e^{-ik_z z} . 
\end{equation}
The field $\tilde{\phi}(x^{(3)},k_z)$ is dual to $\phi(x^{(3)},z)$; these two fields
encode the same information about the system. Indeed, the Fourier transform can be inverted as
\begin{equation}\label{inv-four}
 \phi(x^{(3)},z)=\int\frac{dk_z}{2\pi} \,\tilde{\phi}(x^{(3)},k_z) \, e^{ik_z z} . 
\end{equation}
The reality condition $\phi=\phi^*$ implies
\begin{equation}\label{real}
 \tilde{\phi}(x^{(3)},-k_z)=\tilde{\phi}^*(x^{(3)},k_z).
\end{equation}

Now the idea is to express the action (\ref{unpA}) as a functional of the dual field $\tilde{\phi}$.
By inserting (\ref{inv-four}) into (\ref{unpA}) and recognizing certain integrals as
$\delta$-functions, we obtain
\begin{equation}\label{unpA2}
 A=-\int\frac{dk_z}{2\pi} \, \frac{1}{2}\int d^3x^{(3)}\, \left[
\eta^{ab} \partial_{a}\tilde{\phi}(x^{(3)},-k_z) \, \partial_{b}\tilde{\phi}(x^{(3)},k_z)+
k_z^2 \tilde{\phi}(x^{(3)},-k_z)\tilde{\phi}(x^{(3)},k_z) \right] ,
\end{equation}
where $a,b=0,1,2$.
Using (\ref{real}), this can be written as
\begin{equation}\label{unpA3}
 A=-\int_{-\infty}^{\infty}\frac{dk_z}{2\pi} \, \frac{1}{2}\int d^3x^{(3)}\, \left[
\eta^{ab} \partial_{a}\tilde{\phi}^*(x^{(3)},k_z) \, \partial_{b}\tilde{\phi}(x^{(3)},k_z)+
k_z^2 \tilde{\phi}^*(x^{(3)},k_z)\tilde{\phi}(x^{(3)},k_z) \right] ,
\end{equation}
where we have explicitly written the range of integration over $dk_z$.
Finally, we can introduce a new integration variable $m= |k_z|$
and define
\begin{equation}
\tilde{\phi}(x^{(3)},k_z)\equiv\left\{ 
\begin{array}{c}
 \tilde{\phi}_+(x^{(3)},m) \;\; {\rm for} \;\; k_z>0 , \\
 \tilde{\phi}_-(x^{(3)},m) \;\; {\rm for} \;\; k_z<0 .
\end{array} \right.
\end{equation}
This allows us to write (\ref{unpA3}) as
\begin{equation}\label{unpA4}
 A=\int_{0}^{\infty}\frac{dm}{2\pi} \,[A_+(m)+A_-(m)] ,
\end{equation}
where
\begin{equation}\label{unpA5}
 A_{\pm}(m)=-\frac{1}{2}\int d^3x^{(3)}\, \left[
\eta^{ab} \partial_{a}\tilde{\phi}_{\pm}^*(x^{(3)},m) \, \partial_{b}\tilde{\phi}_{\pm}(x^{(3)},m)+
m^2 \tilde{\phi}_{\pm}^*(x^{(3)},m)\tilde{\phi}_{\pm}(x^{(3)},m) \right] .
\end{equation}

The actions $A_+(m)$ and $A_-(m)$ in (\ref{unpA5}) can be recognized as 3-dimensional actions for a
massive field with mass $m$. These actions are not scale invariant. But we have started
from a scale invariant action (\ref{unpA}), so the action written in terms of the dual 
fields should also be scale invariant. This, indeed, is true because the full action 
(\ref{unpA4}) involves the integration over all possible
values of $m$, so is scale invariant.
The scale invariant field theory which involves the integration over all possible masses can be recognized as
unparticle field theory \cite{step,kras,gaete,nik-unpart}.

Now we can see the onion structure of the action (\ref{unpA4})-(\ref{unpA5}).
The action is 4-dimensional because it involves the integration over 4 coordinates
$x^{(3)}\equiv\{x^0,x^1,x^2\}$ and $m$. The fields depend on all 4 coordinates.
However, the action does not involve derivatives $\partial\tilde{\phi}_{\pm}/\partial m$.
It only involves derivatives $\partial\tilde{\phi}_{\pm}/\partial x^a$ for $a=0,1,2$.
This means that the field interactions propagate only in 3 dimensions. 
An observer made of fields with a fixed value of $m$ can never detect the existence
of fields with other masses. In this sense the dynamics may look 3-dimensional,
despite the fact that there are really 4 dimensions. This is how one dimension may become 
hidden in the dual theory, despite the fact that in the original theory
(\ref{unpA}) all dimensions are clearly present.

\section{The hidden dimension in AdS/CFT}
\label{SEC5}

Now we are finally ready to identify the hidden dimension on the CFT side of AdS/CFT correspondence.
Consider the $D$-dimensional AdS metric
\begin{equation}
 ds^2=\frac{L^2}{z^2}[\eta_{\mu\nu}dx^{\mu}dx^{\nu}+dz^2] ,
\end{equation}
with the boundary at $z=0$ where the $(D-1)$-dimensional CFT theory is supposed to live. 
To avoid the divergence of the metric at the boundary, 
one can consider the theory with a boundary at a small but non-zero $z=\epsilon$. 
This IR cutoff in the AdS spacetime corresponds to a CFT theory with the 
UV cutoff $\Lambda\propto \epsilon^{-1}$ \cite{susskind-witten,peet-polchinski,mcgreevy}.
Thus, in principle, each value of the AdS coordinate $z$ can be associated with a different
CFT theory specified by a different value of the UV cutoff $\Lambda$. So instead of one action
$A_{\rm CFT}$ at $z=0$, we have a continuum of different actions $A_{\rm CFT}(\Lambda)$ at
\begin{equation}\label{zL}
 z\propto\frac{1}{\Lambda} .
\end{equation}

In the standard holographic interpretation of this, 
the parameter $\Lambda$ is viewed merely as a regulator.
Different degrees of freedom are localized at different values of ${\bf x}=\{x^1,\ldots,x^{D-2}\}$,
but different values of $\Lambda$ do not correspond to different local degrees of freedom.
The change of $\Lambda$ is associated with the renormalization-group flow,
which can be related to the change of energy of given degrees of freedom. 
 
The renormalization-group flow in the CFT theory can be related to the dependence on $z$ 
in the bulk theory on AdS \cite{RG1,RG2,RG3}.
However, the idea that different values of $z$ do not correspond to different degrees of freedom
looks somewhat peculiar from the bulk point of view. At the very least, it implies that 
the quantum theory of gravity in the bulk is not a local theory. While it is certainly 
an interesting possibility in a full accordance with the holographic interpretation of
AdS/CFT duality, we believe that it is not the only logical possibility. 
We explore another possibility, the possibility of a non-holographic interpretation. 

According to the non-holographic interpretation of AdS/CFT duality, the theory in the AdS bulk is a local theory,
at least at distances much larger than the Planck length. Different values of {\em both} ${\bf x}$ and $z$
correspond to different degrees of freedom. But we have already seen that different values of $z$ 
correspond to different values of the cutoff $\Lambda$ in the CFT theory, so we see that 
different values of $\Lambda$ should correspond to different degrees of freedom in the CFT theory.
In other words, {\em the non-holographic version of AdS/CFT duality is a conjecture 
that quantum theory of gravity in AdS is dual to the CFT in which different values of ${\bf x}$ and $\Lambda$
correspond to different degrees of freedom}.

Let us make it more precise. Let the CFT theory on the boundary at $z=0$ 
be a renormalizable gauge theory with the action of the form
\begin{equation}\label{A0}
A_{\rm gauge}[\phi,g]=\int d^{D-1}x \,{\cal L}_{\rm gauge}(\phi(x),\partial_{\mu}\phi(x),g) ,   
\end{equation}
where $\phi(x)$ and $g$ are generic fields and coupling constants, respectively. By introducing
a Wilsonian cutoff $\Lambda$ and making the appropriate renormalization (see e.g. \cite{peskin}),
one obtains an action with $\Lambda$-dependent quantities
\begin{equation}\label{AL}
A_{\rm gauge}[\phi(\Lambda),g(\Lambda)]=\int d^{D-1}x \,
{\cal L}_{\rm gauge}(\phi(x,\Lambda),\partial_{\mu}\phi(x,\Lambda),g(\Lambda)) .   
\end{equation}
For instance, $\phi(x,\Lambda)$ can be represented by a Fourier integral of the form
\begin{equation}
 \phi(x,\Lambda)=\int^{\Lambda} \frac{d^{D-1}k}{(2\pi)^{D-1}}\, \tilde{\phi}(k) e^{ik\cdot x} ,
\end{equation}
where $\Lambda$ is the UV cutoff in the momentum space. 

Eq.~(\ref{AL}) is the standard gauge theory expressed in terms of $\Lambda$-dependent quantities,
where different values of $\Lambda$ do not correspond to different degrees of freedom.
It is the gauge theory that appears in the standard holographic AdS/CFT duality. However, in the non-holographic
version different values of $\Lambda$ correspond to different degrees of freedom.
This means that the full action is not (\ref{AL}), but a continuous sum of all such actions
\begin{equation}\label{Af}
 A_{\rm gauge}^{\rm (full)}=\int_0^{\infty} d\Lambda\,  h(\Lambda) \, A_{\rm gauge}[\phi(\Lambda),g(\Lambda)] .
\end{equation}
Here $h(\Lambda)$ is a measure which depends on $\Lambda$ explicitly, but does not depend on
$\phi(\Lambda)$ and $g(\Lambda)$.  
The action (\ref{Af}) has the same symmetries as the action (\ref{AL}), but
has one dimension more corresponding to $\Lambda$. 

The extra dimension parameterized by 
$\Lambda$ is very different from other dimensions parameterized by $x^{\mu}$.
As a consequence, the $D$-dimensional Lorentz group is not a symmetry of (\ref{Af}).
This absence of $D$-dimensional Lorentz symmetry is a way out of the Weinberg-Witten theorem \cite{ww},
the theorem which otherwise would prevent a $D$-dimensional Lorentz-invariant gauge theory to be equivalent
to a $D$-dimensional gravity theory.

The measure $h(\Lambda)$ can be determined from the requirement that the action should be dual 
to the local action in AdS. The full measure in AdS is proportional to
\begin{equation}
 d^Dx \sqrt{|g^{\rm (D)}|}=d^{D-1}x \sqrt{|g^{\rm (D-1)}|}\cdot dz \sqrt{g_{zz}} .
\end{equation}
The measure $d\Lambda\,  h(\Lambda)$ should be dual to the measure $dz \sqrt{g_{zz}}$, so we have
\begin{equation}
 dz \sqrt{g_{zz}}=dz\frac{L}{z}\propto \frac{d\Lambda}{\Lambda} ,
\end{equation}
where we used (\ref{zL}). This implies $h(\Lambda)\propto \Lambda^{-1}$, so (\ref{Af}) can be written as
\begin{equation}\label{Af2}
 A_{\rm gauge}^{\rm (full)}=\int_0^{\infty} \frac{d\Lambda}{c\,\Lambda} \, A_{\rm gauge}[\phi(\Lambda),g(\Lambda)] .
\end{equation}
where $c$ is a constant. The action (\ref{AL}) is dimensionless and 
the full action (\ref{Af2}) must also be dimensionless, so the constant $c$ is dimensionless.
Hence the full action (\ref{Af2}) is determined up to an dimensionless multiplicative constant $c$.

The full action (\ref{Af2}) is very much analogous to the unparticle action (\ref{unpA4}). The integration over 
$d\Lambda$ is analogous to the integration over $dm$, the dimensionless constant $c$ is analogous to the 
dimensionless constant $2\pi$, and the ``ordinary'' action (\ref{AL}) with $\Lambda$-dependent
terms is analogous to the 
``ordinary'' action (\ref{unpA5}) with an $m$-dependent term.  

Similarly to (\ref{unpA4}), we also see the onion structure in the action (\ref{Af2}). The action
(\ref{Af2}) depends on $\partial\phi/\partial x^{\mu}$ (see (\ref{AL})), but it
does not depend on $\partial\phi/\partial\Lambda$. Therefore the interactions propagate in 
$x^{\mu}$-directions but not in the $\Lambda$-direction, which makes the $\Lambda$-dimension hidden.
In this way the theory looks $(D-1)$-dimensional, despite the fact that the full action lives in $D$ dimensions.
That can explain why the standard holographic AdS/CFT correspondence works so well
in many practical cases, even if the full AdS/CFT duality is not fundamentally holographic.
Thus it seems justified to conjecture that $D$-dimensional quantum gravity on AdS might be equivalent to 
the $D$-dimensional action (\ref{Af2}) rather than the $(D-1)$-dimensional action (\ref{A0}). 

\section{Conclusions}
\label{SEC6}

The standard holographic formulation of AdS/CFT duality can explain why the black-hole entropy
is equal to the Bekenstein-Hawking entropy, which is one of the main motivations
for proposing such holographic AdS/CFT duality in the first place.
However, the holographic AdS/CFT duality is not so successful in resolving the 
black-hole information paradox. In this paper we have argued that black-hole information paradox
is much easier to resolve if AdS/CFT duality is modified by a non-holographic version,
in which maximal possible black hole entropy scales with volume rather than area. 
If the non-holographic version is right, then Bekenstein-Hawking entropy can be naturally
reinterpreted as the minimal possible entropy of the black hole, relevant when the entropy of Hawking
radiation can be neglected.

The non-holographic version of AdS/CFT duality is a conjecture that quantum gravity on AdS
and its dual CFT both live in the same number of dimensions.
We have pointed out that it is compatible with the GKPW relation because,
even though this relation expresses a {\em relation} between theories in different numbers of dimensions, 
it does not express a {\em duality} between them. We have supported it by deriving 
a generalized GKPW relation, which is valid for theories which are clearly not dual to each other.

To explain the success of the standard holographic AdS/CFT correspondence we have argued that
one of the dimensions in the dual CFT theory is hidden by the onion structure of the action,
due to which the interactions do not propagate along that dimension. We have identified
this extra dimension to be the UV cutoff $\Lambda$ in the CFT theory. 

We believe that our results make the non-holographic version of AdS/CFT duality a
viable alternative to the standard holographic version. Of course, there is still
no full proof that either of the versions is completely correct, so further research
is certainly needed.

\section*{Acknowledgements}

The author is grateful to D. Jurman for general discussions on AdS/CFT correspondence.
This work was supported by the Ministry of Science of the Republic of Croatia.


\begin{thebibliography}{99}

\bibitem{maldacena}
J.M. Maldacena, Adv. Theor. Math. Phys. {\bf 2}, 231 (1998); hep-th/9711200
\bibitem{gkp}
S.S. Gubser, I.R. Klebanov, A.M. Polyakov, Phys. Lett. B {\bf 428}, 105 (1998); hep-th/9802109.
\bibitem{witten} 
E. Witten, Adv. Theor. Math. Phys. {\bf 2}, 253 (1998); hep-th/9802150.

\bibitem{aharony} 
O. Aharony, S.S. Gubser, J. Maldacena, H. Oogury, Y. Oz,  Phys. Rept. {\bf 323}, 183 (2000); hep-th/9905111.

\bibitem{natsuume} 
M. Natsuume, {\it AdS/CFT Duality User Guide} (Springer, Tokyo, 2015); arXiv:1409.3575.

\bibitem{ammon}
M. Ammon, J. Erdmenger, {\it Gauge/Gravity Duality: Foundations and Applications}
(Cambridge University Press, Cambridge, 2015).

\bibitem{QCD1}
J. Erlich, E. Katz, D.T. Son, M.A. Stephanov, Phys. Rev. Lett. {\bf 95}, 261602 (2005); hep-ph/0501128.
\bibitem{QCD2}
L. Da Rold, A. Pomarol, Nucl. Phys. B {\bf 721}, 79 (2005); hep-ph/0501218.
\bibitem{QCD3}
J. Casalderrey-Solana {\it et al}, {\it Gauge/String Duality, Hot QCD and Heavy Ion Collisions}
(Cambridge University Press, New York, 2014).

\bibitem{fluid1}
G. Policastro, D.T. Son, A.O. Starinets, Phys. Rev. Lett. {\bf 87}, 081601 (2001); hep-th/0104066.
\bibitem{fluid2}
G. Policastro, D.T. Son, A.O. Starinets, JHEP {\bf 0209}, 043 (2002); hep-th/0205052.
\bibitem{fluid3}
M. Rangamani, Class. Quant. Grav. {\bf 26}, 224003 (2009); arXiv:0905.4352.
\bibitem{fluid4}
V.E. Hubeny, S. Minwalla, M. Rangamani, in G.T. Horowitz (ed), {\it Black Holes in Higher Dimensions}
(Cambridge University Press, Cambridge, 2012); arXiv:1107.5780.

\bibitem{supercond1}
S.A. Hartnoll, C.P. Herzog, G.T. Horowitz, Phys. Rev. Lett. {\bf 101}, 031601 (2008); arXiv:0803.3295.
\bibitem{supercond2}
S.A. Hartnoll, C.P. Herzog, G.T. Horowitz, JHEP {\bf 0812}, 015 (2008); arXiv:0810.1563.
\bibitem{supercond3}
C.P. Herzog, J. Phys. A {\bf 42}, 343001 (2009); arXiv:0904.1975.

\bibitem{ent1}
S. Ryu, T. Takayanagi, 
Phys. Rev. Lett. {\bf 96}, 181602 (2006); hep-th/0603001.
\bibitem{ent2}
S. Ryu, T. Takayanagi, 
JHEP {\bf 0608}, 045 (2006); hep-th/0605073.
\bibitem{ent3}
T. Nishioka, S. Ryu, T. Takayanagi, J. Phys. A {\bf 42}, 504008 (2009); arXiv:0905.0932.


\bibitem{bekenstein}
J.D. Bekenstein, Phys. Rev. D {\bf 7}, 2333 (1973).
\bibitem{bch}
J.M. Bardeen, B. Carter, S.W. Hawking, Commun. Math. Phys. {\bf 31}, 161 (1973).

\bibitem{BD}
N.D. Birrell, P.C.W. Davies, {\it Quantum Fields in Curved Space} (Cambridge Press, NY, 1982).

\bibitem{hawking}
S.W.~Hawking, Commun.~Math.~Phys.~{\bf 43}, 199 (1975).

\bibitem{gid}
S.B. Giddings, Phys. Rev. D {\bf 46}, 1347 (1992); hep-th/9203059. 
\bibitem{har} 
J.A.~Harvey, A.~Strominger, hep-th/9209055.
\bibitem{pres}
J. Preskill, hep-th/9209058.
\bibitem{pag} 
D.N.~Page, hep-th/9305040.
\bibitem{gid2} 
S.B.~Giddings, hep-th/9412138.
\bibitem{str}
A.~Strominger, hep-th/9501071.
\bibitem{math1}
S.D. Mathur, Lect. Notes Phys. {\bf 769}, 3 (2009); arXiv:0803.2030.
\bibitem{math2}
S.D. Mathur, Class. Quant. Grav. {\bf 26}, 224001 (2009); arXiv:0909.1038.
\bibitem{hoss}
S. Hossenfelder, L. Smolin, Phys. Rev. D {\bf 81}, 064009 (2010); arXiv:0901.3156.
\bibitem{harlow}
D. Harlow, arXiv:1409.1231.
\bibitem{fabbri}
A. Fabbri, J. Navarro-Salas, {\it Modeling Black Hole Evaporation}
(Imperial College Press, London, 2005).


\bibitem{nik1}
H. Nikoli\'c, Phys. Lett. B {\bf 678}, 218 (2009); arXiv:0905.0538.
\bibitem{nik2}
H. Nikoli\'c,  Int. J. Quantum Inf. {\bf 10}, 1250024 (2012); arXiv:0912.1938.
\bibitem{nik3}
H. Nikoli\'c,  Int. J. Quantum Inf. {\bf 12}, 1560001 (2014); arXiv:1407.8058.
\bibitem{nik4}
H. Nikoli\'c, JCAP {\bf 04}, 002 (2015); arXiv:1502.04324.

\bibitem{nik-crystal}
H. Nikoli\'c, arXiv:1505.04088.

\bibitem{strominger-vafa}
A. Strominger, C. Vafa, Phys. Lett. B {\bf 379}, 99 (1996); hep-th/9601029.
\bibitem{horowitz}
G.T. Horowitz, gr-qc/9704072.
\bibitem{peet}
A.W. Peet, hep-th/0008241.

\bibitem{hsu}
S.D.H. Hsu, D. Reeb, Phys. Lett. B {\bf 658}, 244 (2008); arXiv:0706.3239.

\bibitem{mathur-dialog}
S.D. Mathur, {\it Confusions and questions about the information paradox},
http://www.physics.ohio-state.edu/\verb|~|mathur/confusions2.pdf

\bibitem{bek-bound}
J.D. Bekenstein, Phys. Rev. D {\bf 23}, 287 (1981).

\bibitem{susskind-lindesay}
L. Susskind, J. Lindesay, {\it An Introduction to Black Holes, Information, and the String Theory Revolution:
The Holographic Universe} (World Scientific Publishing, Singapore, 2005). 

\bibitem{smolin-bek-bound}
L. Smolin, Nucl. Phys. B {\bf 601}, 209 (2001); hep-th/0003056.

\bibitem{georgi}
H. Georgi, Phys.~Rev.~Lett.~{\bf 98}, 221601 (2007); hep-ph/0703260.
 
\bibitem{step}
M.A. Stephanov, Phys.~Rev.~D {\bf 76}, 035008 (2007); arXiv:0705.3049.
\bibitem{kras}
N.V.~Krasnikov, Int.~J.~Mod.~Phys.~A {\bf 22}, 5117 (2007); arXiv:0707.1419.
\bibitem{gaete}
P.~Gaete, E.~Spallucci, Phys. Lett. B {\bf 661}, 319 (2008); arXiv:0801.2294.
\bibitem{nik-unpart}
H. Nikoli\'c, Mod. Phys. Lett. A {\bf 23}, 2645 (2008); arXiv:0801.4471.

\bibitem{cacci}
G. Cacciapaglia, G. Marandella, J. Terning, JHEP {\bf 0902}, 049 (2009); arXiv:0804.0424.

\bibitem{susskind-witten}
L. Susskind, E. Witten, hep-th/9805114.
\bibitem{peet-polchinski}
A.W. Peet, J. Polchinski, Phys. Rev. D {\bf 59}, 065011 (1999); hep-th/9809022.
\bibitem{mcgreevy}
J. McGreevy, Adv. High Energy Phys. 2010, 723105 (2010); arXiv:0909.0518.

\bibitem{RG1}
E.T. Akhmedov, Phys. Lett. B {\bf 442}, (1998); hep-th/9806217.
\bibitem{RG2}
M. Porrati, A. Starinets, Phys. Lett. B {\bf 454}, 77 (1999); hep-th/9903085.
\bibitem{RG3}
V. Balasubramanian, P. Kraus, Phys. Rev. Lett. {\bf 83}, 3605 (1999); hep-th/9903190.

\bibitem{peskin}
M.E. Peskin, D.V. Schroeder, {\it An Introduction to Quantum Field Theory}
(Perseus Books, Massachusetts, 1995).

\bibitem{ww}
S. Weinberg, E. Witten, Phys. Lett. B {\bf 96}, 59 (1980).  


\end{thebibliography}
\end{document}